\documentclass[aps,prl,twocolumn,groupedaddress,showpacs]{revtex4}

\usepackage{amsmath}
\usepackage{graphicx}

\newcommand{\nn}{\nonumber}
\newcommand{\ob}[1]{\overline{#1}}

\newcommand{\aou}{\hat{a}}
\newcommand{\aod}{\hat{a}^{\dagger}}
\newcommand{\ai}{\hat{a}}
\newcommand{\aid}{\hat{a}^{\dagger}}

\newcommand{\n}{\hat{n}}
\renewcommand{\ni}{\n}

\newcommand{\bt}[1]{\langle#1\rangle}

\begin{document}

\title{Quantum interference and entanglement induced by multiple scattering of light}


\author{J.~R. \surname{Ott}}\email{jrot@fotonik.dtu.dk}
\author{N.~A. \surname{Mortensen}}
\author{P. \surname{Lodahl}}
\affiliation{Department of Photonics Engineering, Technical University of Denmark,  DK-2800 Kgs. Lyngby, Denmark}

\date{\today}

\begin{abstract}
We report on the effects of quantum interference induced by transmission of an arbitrary number of optical quantum states through a multiple scattering medium. We identify the role of quantum interference on the photon correlations and the degree of continuous variable entanglement between two output modes. It is shown that the effect of quantum interference survives averaging over all ensembles of disorder and manifests itself as increased photon correlations giving rise to photon anti-bunching. Finally, the existence of continuous variable entanglement correlations in a volume speckle pattern is predicted. Our results suggest that multiple scattering provides a promising way of coherently interfering many independent quantum states of light of potential use in quantum information processing.
\end{abstract}

\pacs{42.25.Dd, 42.50.Lc, 78.67.-n}

\maketitle
Studies of wave propagation in disordered media have revealed a range of fascinating wave phenomena, including Anderson localization~\cite{anderson_pr58a}, enhanced coherent back scattering~\cite{albada_prl85a_wolf_prl85a}, and universal conductance fluctuations~\cite{Lee:1985}. These phenomena, originating from wave interference, appear after averaging over all configurations of disorder, in the mesocopic regime~\cite{rossum_rmp99a,beenakker_rmp97a}. Mesoscopic effects are common to a large class of elastic wave problems, including transport of electrons through artificial nanostructures, advancement of sound waves in turbulent fluids, and light propagation through complex dielectric media.

While much attention has been devoted to propagation of classical light waves through random media over the years, the influence of the quantum nature of light has only recently become an active field of research. The development of a theoretical framework to handle multiple scattering in quantum optics~\cite{beenakker_prl98a,patra_pra00a} triggered the interest in understanding quantum optical properties of disordered media~\cite{beenakker_prl98a, lodahl_prl05b,patra_pra00a, tworzydlo_prl02a, aiello_pra04a, velsen_pra04a,   lodahl_prl05a_lodahl_oe06a, smolka_prl09a,beenakker_prl09a,lahini_arxiv10a,peeters_prl10a}. Studies include quantum noise properties~\cite{beenakker_prl98a,lodahl_prl05b}, absorbing or amplifying media~\cite{patra_pra00a,tworzydlo_prl02a}, degradation of polarization entanglement~\cite{aiello_pra04a},
 and spatial photon correlations~\cite{lodahl_prl05a_lodahl_oe06a, smolka_prl09a,beenakker_prl09a,lahini_arxiv10a,peeters_prl10a}. Recently it was shown experimentally that light-matter interaction is strongly enhanced in disordered photonic crystal waveguides, enabling cavity quantum electrodynamics with Anderson-localized modes~\cite{sapienza_science10a}. Optical quantum information processing schemes rely on interference among multiple independent quantum states, i.e. quantum interference (QI), to generate quantum correlations and entanglement. The possibility of using multiple-scattering media to interfere independent quantum states is appealing since it is inherently scalable to multiple input states. To this end, mesoscopic quantum interference effects that persist even after averaging over all ensembles of disorder, would be required in order to obtain robust and predictable quantum correlations. So far, interference of quantum light in a scattering medium has only been described in 1D random walk models~\cite{lahini_arxiv10a} or for diffusive transport where interference effects wash out after ensemble averaging~\cite{beenakker_prl09a,peeters_prl10a}.
\begin{figure}[t!]
	\centering
		\includegraphics[width=1\columnwidth]{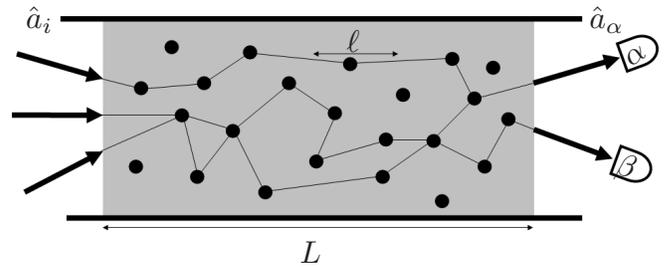}
	\caption{(color online). Sketch of propagation through a disordered waveguide of length $L$ and transport mean free path $\ell$. Quantized light is incident to the left and the correlations between two output modes on the right are analyzed. The operators $\ai_i$ and $\aou_\alpha$ corresponds to the creation operators of modes $i$ and $\alpha$, where Roman and Greek subscripts denote input and output modes respectively. The correlations between the two different output modes $\alpha$ and $\beta$ are analyzed.}
	\label{fig:1}
\end{figure}

In this Letter, we investigate QI induced by combining an arbitrary number of independent quantum states in a random multiple scattering medium in the mesoscopic regime. We identify the role of QI on the degree of photon number correlations between two transmission paths through the medium and the degree of continuous variable entanglement. Surprisingly QI of photons is found to survive after averaging over all configurations of disorder in the mesoscopic regime, i.e. the induced quantum correlations have deterministic character despite the underlying random multiple scattering processes. At last we discuss the feasibility of experimentally verifying our theoretical predictions.

Let us now introduce the model for propagation of quantized light through a linear, elastic, multiple scattering medium of length $L$ and transport mean free path $\ell$, see Fig.~\ref{fig:1}. We apply the scattering matrix for the propagation of light and use random matrix theory on the scattering elements. The approach describes effectively a quasi-1D model of an $N$-mode waveguide, but is known also to accurately predict propagation in 3D slab geometries~\cite{beenakker_rmp97a}. We relate the photon annihilation operators $\hat{a}_{\alpha}$ $(\hat{a}_{i})$ of output (input) modes $\alpha$ $(i)$ by $\hat{a}_{\alpha}$$=$$\sum_it_{\alpha i}\hat{a}_i$, where the summation is over all $N$ possible input modes at each end of the waveguide and $t_{\alpha i}$ denotes the complex scattering matrix element. Experimentally such a system could, e.g., be realized in titania powder samples~\cite{smolka_prl09a} or disordered photonic crystal waveguides~\cite{sapienza_science10a,smolka_in_prep_10a}.

As a measure of QI, we introduce the 2-channel photon correlation function
\begin{align}
C_{\alpha\beta}=\frac{\Delta \ni_\alpha\ni_\beta}{\bt{\ni_\alpha}\bt{\ni_\beta}},\label{eqn:g2_2}
\end{align}
where
\begin{align}
\Delta \ni_\alpha\ni_\beta=\bt{\ni_\alpha\ni_\beta}-\bt{\ni_\alpha}\bt{\ni_\beta}.\label{eqn:g2}
\end{align}
The brackets denote quantum mechanical expectation values and $\ni_\alpha$$=$$\aod_\alpha\aou_\alpha$ is the output photon number operator. The degree of entanglement is quantified in terms of the quadrature variance product (QVP)
\begin{align}
\varepsilon_{\alpha\beta}=\Delta (\hat{X}_\alpha-\hat{X}_\beta)^2\Delta (\hat{Y}_{\alpha}+\hat{Y}_{\beta})^2,\label{eqn:EPR}
\end{align}
where $\hat{X}_\alpha$$=$$\frac{1}{\sqrt{2}}(\aod_\alpha$$+$$\aou_\alpha)$ and $\hat{Y}_\alpha$$=$$\frac{i}{\sqrt{2}}(\aod_\alpha$$-$$\aou_\alpha)$ are quadrature operators.  The QVP determines the ability to predict a measurement in mode $\beta$ given the result of a measurement on mode $\alpha$, and for $\varepsilon_{\alpha\beta}$$<$$1$ ($>$$1$) the outcome is predictable below (above) the quantum noise limit. Here, $\varepsilon_{\alpha\beta}$$<$$1$ implies that the quantum state of the two output modes $\alpha$ and $\beta$ is unseparable, i.e. entangled~\cite{mancini_prl02a}.
\begin{figure}[t!]
	\centering
		\includegraphics[width=1\columnwidth]{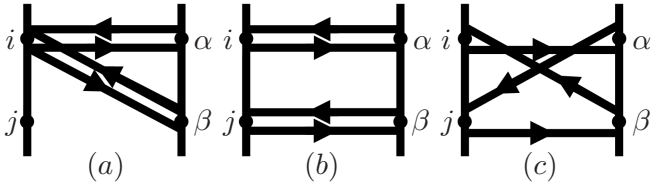}
	\caption{A diagrammatic representation of three terms of the form $t_{\alpha i}^*t_{\beta j}^*t_{\beta k}t_{\alpha l}\bt{\aid_i\aid_j\ai_k\ai_l}$. Diagrams (a) and (b) involve only intensity transmission of the input modes while (c) shows quantum interference between two input states. The three diagrams are the only ones that survive ensemble averaging.}
	\label{fig:2}
\end{figure}
Eqs.~\eqref{eqn:g2_2} to~\eqref{eqn:EPR} can conveniently be evaluated diagrammatically, by representing the propagator $t_{\alpha i}\ai_i$ by an arrow connecting input mode $i$ to output mode $\alpha$. Since the scattering matrix is unitary, $t_{\alpha i}^*\aid_i$ represents the time-reversed path. E.g. considering two input modes, evaluating Eq.~\eqref{eqn:g2} yields $2^4$ different terms of the form $t_{\alpha i}^*t_{\beta j}^*t_{\beta k}t_{\alpha l}\bt{\aid_i\aid_j\ai_k\ai_l}$. Three typical diagrams are shown in Fig.~\ref{fig:2}. As an example, the contribution from diagram (a) is $|t_{\alpha i}|^2|t_{\beta i}|^2\bt{:\ni_i^2:}$, where $\bt{: \cdot :}$ denotes normal ordering and $\ni_i$$=$$\aid_i\ai_i$ is the input photon number operator. The diagrams can be classified into intensity and interference diagrams. The former is an incoherent addition of the intensities associated with the different propagation paths through the medium, as it is the case for the diagrams~(a) and~(b). The latter gives rise to QI between the input states. Diagram~(c) is an example that show such interference. We note that all additional diagrams not shown in Fig.~\ref{fig:2} are interference diagrams.

\begin{figure}[b!]
	\centering
		\includegraphics[width=0.9\columnwidth]{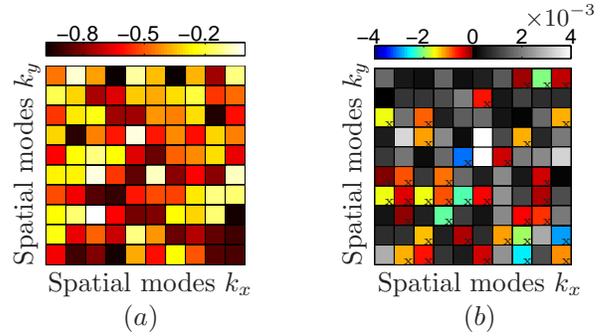}
	\caption{(color online). (a) The 2-photon correlation $C_{\alpha\beta}$ for two single-photon Fock input states, $\left|1,1 \right>$, showing large fluctuations due to quantum interference. (b) The degree of entanglement $\log_{10}(\varepsilon_{\alpha\beta})$ for the two quadrature squeezed input states described in the text, where the gray-scale corresponds to non-entangled states $(\log_{10}(\varepsilon_{\alpha\beta})$$\ge$$0)$ while the colored areas with crosses display the entangled states $(\log_{10}(\varepsilon_{\alpha\beta})$$<$$0)$. (a) and (b) are obtained with the same realization of disorder for diffusive transport, where the phase is random and $P(|t_{\alpha i}|^2)$$=$$\exp(-|t_{\alpha i}|^2/\ob{|t_{\alpha i}|^2})$ with $\ob{|t_{\alpha i}|^2}$$=$$\tau$ the average single-channel transmission.}
	\label{fig:3}
\end{figure}

For diffusive transport the intensity transmission coefficients are exponentially distributed while the phase is uniformly distributed~\cite{nieuwenhuizen_prl95a}. Consequently, this allows us to generate a set of normalized amplitude transmission coefficients valid for a single realization of disorder. To be specific, we choose a waveguide of $N$$=$$10^2$ so that the average single channel transmission is $\tau$$=$$g/N^2$$=$$1/300$~\cite{footnote1}, where $g$ is the normalized average conductance. The modes are written in a 10 by 10 grid to represent spatial wavevectors in the transverse plane ($k_x$, $k_y$). We evaluate the 2-channel photon correlation function using Fock states $\left|n \right>$ as input. Illuminating only a single input channel with a two-photon Fock state $\left|2 \right>$, $C_{\alpha\beta}$$=$$-1/2$ for all modes $\alpha$ and $\beta$, independent of the realization of disorder. For single photons incident in two different input modes $\left|1, 1\right>$ the spatial photon correlations fluctuate between -1 and 0, see Fig.~\ref{fig:3}~(a). This is a manifestation of QI in a speckle pattern as observed in Ref.~\cite{peeters_prl10a}.
We furthermore evaluate $\varepsilon_{\alpha\beta}$ for two quadrature-squeezed input states $\left|\zeta_i\right>$$=$$\exp[\frac{1}{2}\zeta_i^*\ai^2-\frac{1}{2}\zeta_i(\aid)^2] \left|0\right>$ where $\zeta_i$$=$$|\zeta_i|e^{i\phi_i}$ contains squeezing amplitude, $|\zeta|$, and phases, $\phi_i$,~\cite{RLoudon_3rd_ed2007}. We investigate two orthogonally oriented squeezed beams, i.e. $\phi_i$$=$$0$ and $\phi_j$$=$$\pi$ and chose $|\zeta|$$=$$0.15$  corresponding to experimentally obtainable parameters~\cite{smolka_prl09a}. Fig.~\ref{fig:3}~(b) is a calculation of $\varepsilon_{\alpha\beta}$ for different output modes and show that entanglement ($\varepsilon_{\alpha\beta}$$<$$1$) can be induced by multiple scattering. If we change the squeezing phases of the input states the modes that display entanglement change. From knowledge of the transmission matrix one could thus specify the mode in which the entanglement should occur by changing the phases of the squeezing parameters. This could potentially be achieved with the recent scheme to measure the complex transmission matrix for light propagation through a disordered medium~\cite{popoff_prl10a}. The inherent ability of a multiple scattering medium to mix many modes shows the scalability of the approach of potential use in quantum information processing.

Let us next consider the effects of QI after ensemble averaging. For this we need the averaged amplitude transmission coefficients which are given by~\cite{cwilich_pre06a}
\begin{subequations}
\begin{align}
\ob{t_{\alpha i}^*t_{\alpha j}}&=\tau\delta_{ij},\\
\ob{t_{\alpha i}^*t_{\beta j}^*t_{\beta k}t_{\alpha l}}&=\tau^2\left(C_1\delta_{il}\delta_{jk}+C_2\delta_{ik}\delta_{jl}\right),
\end{align}
\end{subequations}
with $C_1$ and $C_2$ the short and long-range correlation functions respectively and the bar denoting ensemble averaging. When carrying out the ensemble average only loop-type diagrams survive, i.e. only those shown in Fig.~\ref{fig:2}. The values of diagrams (a) and (b) are proportional to $C_1$$+$$C_2$ and $C_1$ respectively while diagram (c) is proportional to $C_2$. This can be intuitively understood since $C_1$ expresses the intensity fluctuations in a single speckle spot, while $C_2$ is related to the intensity correlations between two speckles~\cite{garcia_martin_prl02a}. Diagram (c) displays the QI term that survives ensemble averaging. From the values of $C_2$ and the normalized average conductance $g$, we define the transitions from the quasi-ballistic to the weakly disordered regime ($C_2$$=$$0$) and from the weakly disordered to the localized regime ($g$$=1$). The mesoscopic regime is defined as the regime in which two speckle spots are correlated after ensemble averaging ($C_2$$>$$0$). The values of $C_1$ and $C_2$ depend on the number of modes $N$ and the degree of disorder, which is contained in $s$$=$$L/\ell$ and $g^{-1}$~\cite{footnote2}.

We define the ensemble-averaged 2-channel correlation function as
\begin{align}
\ob{C}_{\alpha\beta}\!&=\! \frac{\ob{\Delta \ni_\alpha\ni_\beta}}{\ob{\bt{\ni_\alpha}\bt{\ni_\beta}}}\label{obC}\\
\!&=\!\frac{(C_1\!+\!C_2)\!\!\left[\left({\displaystyle\sum_{i}}\bt{\ni_i}\!\right)^2\!+{\displaystyle\sum_i}(\Delta \ni_i^2-\bt{\ni_i})\right]}{ C_1\!\!\left({\displaystyle\sum_i}\bt{\ni_i}\!\right)^2\!\!+C_2\!\!\left(\!{\displaystyle\sum_i\bt{\ni_i}^2}\!+\!2\!{\displaystyle\sum_{i,j> i}\!|\bt{\aid_i \ai_j}|^2}\!\right)}\!-\!1.\nn
\end{align}
The ensemble averaged QVP is
\begin{align}
\ob{\varepsilon}_{\alpha\beta}=&1+4\tau\!\sum_i\Delta\aid_i\ai_i\\
&+4\tau^2\!\left[C_1\!\left(\sum_i\Delta\aid_i\ai_i\right)^2\!\!+C_2\!\sum_{i,j}\Delta\aid_i\ai_j\Delta\aid_j\ai_i\right].\nn
\end{align}
\begin{figure}[t!]
	\centering
		\includegraphics[width=0.9\columnwidth]{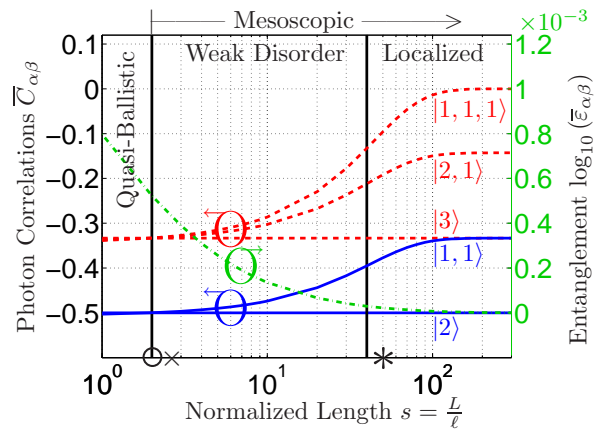}
	\caption{(color online). The ensemble averaged 2-channel photon correlations, $\ob{C}_{\alpha\beta}$ and the degree of entanglement $\log_{10}(\ob{\varepsilon}_{\alpha\beta})$ versus $s$$=$$L/\ell$ for $N=50$. Solid blue curves show $\ob{C}_{\alpha\beta}$ for Fock input states with a total of two photons and dashed red curves are for three photons. The difference in $\ob{C}_{\alpha\beta}$ between having one and more input states is due to quantum interference. The green dash-dotted line shows $\log_{10}(\ob{\varepsilon}_{\alpha\beta})$ for the two quadrature squeezed input states described in the text. The vertical line at $s$$=$$2$ indicates the crossover from the quasi-ballistic to the mesoscopic regime $(C_2$$=$$0)$ and the one at $s$$\approx$$40$ shows the transition to the localized regime $(g$$=$$1)$. The symbols $\circ$, $\times$, and $\ast$ on the abscissa correspond to the experimental structures studied in  Refs.~\cite{peeters_prl10a},~\cite{smolka_prl09a}, and~\cite{smolka_in_prep_10a}, as discussed further in the main text.}
	\label{fig:4}
\end{figure}
In Fig.~\ref{fig:4} $\ob{C}_{\alpha\beta}$ and $\log_{10}(\ob{\varepsilon}_{\alpha\beta})$ are plotted versus $s$. For Fock input states the $s$ dependence of $\ob{C}_{\alpha\beta}$ is a direct measure of QI since disregarding the QI terms implies that $\ob{C}_{\alpha\beta}$ only depends on the total number of input photons. First consider having two photons incident in only one mode, $\left|2\right>$, $\ob{C}_{\alpha\beta}$$=$$-\frac{1}{2}$ independent of $s$. With the photons in two different input modes, $\left|1,1\right>$, the correlations on the contrary depend on $s$. Only at the transition to the mesoscopic regime we have $\ob{C}_{\alpha\beta}$$=$$-\frac{1}{2}$. This value corresponds to the correlation between two equally probable output modes for two classical non-interacting particles. This is due to two simple reasons: (i) the ensemble averaging makes all output modes equally possible and (ii) in this limit transport is diffusive and thus all interference effects are washed out. As disorder is increased, $\ob{C}_{\alpha\beta}$ increases signifying that the probability that the two photons arrive at two different positions increases although remaining anti-correlated $(\ob{C}_{\alpha\beta}$$<$$0)$. The increased correlations saturate in the localized regime since $C_2$ tends towards $C_1$~\cite{garcia_martin_prl02a}. The variations in $\ob{C}_{\alpha\beta}$ can be attributed to QI amongst the input channels, which causes the photons to antibunch. For three photon Fock states the more incident modes the larger the effect of QI on $\ob{C}_{\alpha\beta}$. For $\left|1,1,1\right>$, $\ob{C}_{\alpha\beta}$ tend towards zero in the localized regime signifying that the output modes become uncorrelated. If the number of input modes is increased further the output modes will become correlated ($\ob{C}_{\alpha\beta}$$>$$0$) in the localized regime. This means that detection of a photon in one mode on average increases the probability of detection of a photon in another mode, which is in striking contrast to the behavior of diffusive transport. Having single-photon states in $n$ input modes and letting $n$ go to infinity makes $\ob{C}_{\alpha\beta}$ approach unity far into the localized regime, which is the value obtained for thermal light. The variation of $\ob{\varepsilon}_{\alpha\beta}$ with $s$ is plotted as the green curve in Fig.~\ref{fig:4} for the same quadrature-squeezed inputs as for the single realization of Fig.~\ref{fig:3}. In the mesoscopic regime both $C_1$ and $C_2$ are positive and thus $\ob{\varepsilon}_{\alpha\beta}$$\geq$$1$. The value of $\ob{\varepsilon}_{\alpha\beta}$ approaches unity in the localized regime since the transmission decreases so that contributions from vacuum fluctuations dominate. Continuous variable entanglement in the transmission is therefore predicted to vanish after ensemble averaging. We anticipate that this might be different in the reflection due to coherent effects as enhanced backscattering, but this is outside the scope of the present work.

Finally, we address the experimental feasibility of the proposal. In Fig. \ref{fig:4} we indicate the position of three existing multiple scattering structures from the literature, where the number of modes $N$ has been scaled to match the value used in the calculations. Ref.~\cite{peeters_prl10a} concerns transmission through two scattering surfaces, which mimic a multiple scattering medium with  $s=2$. This corresponds to the diffusive limit where QI will be present in the speckle pattern but not survive ensemble averaging. In Ref.~\cite{smolka_prl09a} a titania powder is used with sample length $L$$=$$20\,\mu$m and transport mean free path $\ell$$\approx$$0.9\,\mu$m, which corresponds to the mesoscopic regime of  $s$$>$$2$. Such sample support a large number of modes ($N$$>$$10^3$) and thus $g$$\gg$$1$, which means that this type of sample is in the weakly disordered regime where QI effects are modest, cf. Fig. \ref{fig:4}. This illustrates the importance of using multiple scattering samples supporting only few modes in order to observe QI. A disordered multimode photonic crystal waveguide is exactly such a system and for $N$$\approx$$5$ together with the typical experimental parameters of $\ell$$\approx$$20\,\mu$m and $L$$=$$100\,\mu$m~\cite{smolka_in_prep_10a} gives rise to sizeable QI effects that will be observable in an experiment, cf. Fig. \ref{fig:4}.

In conclusion we have theoretically predicted that surprisingly QI of light survives multiple scattering even after ensemble averaging and can be employed to mix several input quantum states of light. By calculating the 2-channel photon correlation function in the case of Fock input states, we showed that the difference between one and several input modes for a fixed number of total photons is a direct measure of QI. The effect was found to give rise to spatial photon anti-bunching in the mesoscopic regime and to increase with the degree of disorder. The experimental feasibility of the proposal was investigated based on existing multiple scattering samples from the literature, and we found that multimode disordered photonic crystal waveguides are promising candidates for an experimental demonstration. We furthermore investigated continuous variable entanglement induced by multiple scattering of squeezed light that was predicted to show up in single realizations of disorder but to vanish after ensemble averaging. Our work may provide a promising new route to coherently combine many independent quantum states of light that is inherently scalable since the multiple-scattering process mixes all input states.

The authors acknowledge M. Wubs, S. Smolka, and J. G. Pedersen for comments on the manuscript, U. L. Andersen  for stimulating discussions, and L. S. Froufe-P{\'e}rez for providing the data for $C_1$, $C_2$, and $g$. We gratefully acknowledge the Council for Independent Research (Technology and Production Sciences and Natural Sciences) for financial support.

\begin{thebibliography}{26}
\bibitem{footnote1} For diffusive transport the average single channel transmission is $g$$\approx$$ N/(1+L/\ell)$~\cite{beenakker_rmp97a} and $s$$=$$L/\ell\approx2$~\cite{garcia_martin_prl02a} giving an average amplitude transmission of $\tau$$=$$1/300$ for $N$$=$$10^2$.
\bibitem{footnote2} The values of $C_1$ and $C_2$ are independent of $N$ on the transition between the quasi-ballistic and the mesoscopic regimes, $s$$\approx$$2$, and tend toward the same value far into the localized regime, $g^{-1}$$\ll$$1$~\cite{garcia_martin_prl02a}. The values of $C_1$ and $C_2$ has qualitatively the same behaviors versus disorder in the various regimes for different $N$.
\end{thebibliography}

\end{document}